\documentclass[prb,preprint,a4paper,showpacs]{revtex4}
\usepackage{amsmath}
\usepackage{epsfig}
\begin{document}
\titlepage
\title{\bf S-Matrix Formulation of Mesoscopic Systems and Evanescent Modes}
\author{Sheelan Sengupta Chowdhury$^1$, P. Singha Deo$^1$, 
A. M. Jayannavar$^2$ and M. Manninen$^3$}
\affiliation{$^1$Unit for Nanoscience and Technology, S. N. Bose National 
Centre for Basic Sciences, Sector-III, Block-JD, Salt Lake, Kolkata-700098, 
India\\
$^2$ Institute of Physics, Sachivalaya marg, Bhubaneswar 750015, India\\
$^3$ Nanoscience center, Department of Physics, University of Jyvaskyla,
PO Box 35, 40100 Jyvaskyla, Finland}
\begin{abstract}
The Landauer-Butikker formalism is an important formalism to study
mesoscopic systems. Its validity for linear transport is well established
theoretically as well as experimentally. Akkermans et al 
[Phys. Rev. Lett. {\bf 66}, 76 (1991)] had shown that
the formalism can be extended to study thermodynamic properties like
persistent currents. It was earlier verified for simple one dimensional
systems. We study this formula very carefully and conclude that it requires
reinterpretation in quasi one dimension. This is essentially because
of the presence of evanescent modes in quasi one dimension.
\end{abstract}
\maketitle

\section{Introduction}

Due to the technological advances in nano-fabrication, it is possible
to realize such small systems that the quantum mechanical coherence length
of the electron extends through out the length of the sample.
Quantum interference
phenomena strongly determines the thermodynamic and transport properties of
these so called mesoscopic systems \cite{dat}. 

Mesoscopic phenomenon can occur in canonical systems as well as in grand
canonical systems. 
Mesoscopic systems are so small that even measuring probes (like voltage
probe and current probe) can make the system a grand canonical system
\cite{but86}.
A canonical mesoscopic system is well described by the
Hamiltonian of the isolated
system but that is not the case for grand canonical
mesoscopic systems. A grand canonical mesoscopic system, by definition, is
coupled to a reservoir with which it can exchange electrons. Mesoscopic
systems are so small that the reservoir can drastically change the states
of the system and this has to be explicitly accounted for \cite{but85}. 
One can take this into account by solving the Schrodinger equation
of the leads and the system as a scattering problem. This approach
is essentially known as Landauer-Butikker formalism. This formalism
is thus different from the way we deal with large grand-canonical
systems with the help of the grand partition function.

For example, if an Aharonov Bohm flux
is applied through the
center of a ring, the ring gets magnetized and a persistent current is
generated in the ring \cite{per}. 
This current arises because of the vector potential
that changes the phase of the wave function in the ring and is another form
of Aharonov Bohm effect which is an interference phenomenon, whereas
the magnetization is a thermodynamic property is a consequence of that.
One can cite many similar phenomenon \cite{dat}.
Persistent currents has been studied for more than a decade, theoretically as
well as experimentally.
If the ring is isolated then the persistent current is carried
in some eigen-states. Whereas if it is open and connected to
reservoirs, then persistent current is carried in resonant and
non-resonant states that are typical of scattering states.
Persistent current has been studied in several such grand
canonical systems like a ring connected to a single reservoir or to many
reservoirs \cite{njp}.
If it is connected to many reservoirs at different chemical potentials,
then non-equilibrium currents can co-exist with equilibrium persistent
currents. 
In such open systems several interesting effects have been
predicted, like current magnification in the presence of transport
\cite{am1, am2, am3}, directional dependence of 
persistent currents \cite{am4},
current magnification effect in equilibrium systems in absence of
transport current \cite{am5}, etc.

In order to realize a mesoscopic grand canonical system
we connect the ring to reservoirs that are at fixed chemical potentials
as is schematically shown in Fig.
\ref{figsys}. The left reservoir has a chemical potential $\mu_1$ and the right
one has a chemical potential $\mu_2$. 
The reservoirs can also be at a finite temperature T.
The ring is threaded by an Aharonov-Bohm
flux. If $\mu_1 >\mu_2$, then there is a transport current (which is a
non-equilibrium current) through the regions I
and II. There is however no transport current in the ring. 
The ring will carry a
persistent current which is an equilibrium current. Thus in the present
geometry, the equilibrium persistent currents and non-equilibrium transport
currents are spatially separated.

The Landauer-Buttiker approach proposes that an 
equilibrium phenomenon like
persistent current in such an open system as that
in Fig. 1, can be obtained from solving the scattering problem
\cite{but85}.
Akkermans et al \cite{akk} related the persistent current $IS$ to
the $S$ matrix by the following formula.
\begin{eqnarray}
IS=\frac{1}{2\pi i} \frac{\partial \log[\det(S)]}{\partial \phi}
\label{ja}
\end{eqnarray}
Such a simple mathematical relation between the persistent current
inside the ring and the S-matrix obtained from the wave-function
far away from the ring is rather novel and resulted in a flurry
of theoretical activities \cite{mon}.
While it is established (from theoretical and experimental point of view)
that conductance (a non-equilibrium phenomenon) can be obtained from the
S-matrix, Akkermans approach may prove to be the first step to obtain
any equilibrium phenomenon from the S-matrix, that is a step towards
a mesoscopic version of fluctuation dissipation theorem \cite{mon}.
The correctness of Eq. \ref{ja} has been explicitely verified in 
one dimension (1D)
but not in quasi one dimension (Q1D).
That complexities arise in Q1D due to the presence of evanescent modes
has been observed recently \cite{swa,lar}
although it is not very well known to
the community \cite{aba}. 
For example Friedel sum rule \cite{swa} and Buttiker-Thomas-Pertre formula
\cite{lar} breaks down in the presence of evanescent modes.
So in this work we undertake the task of verifying
if Akkermans formula is valid in Q1D.
In case of Akkerman's formula, in this paper, we can show analytically
how the evanescent modes complicates things. Earlier works 
\cite{swa,lar} on different
formulas are essentially numerical verifications.
Persistent current in the geometry of Fig. 1
has been studied earlier, but always
using the wave function. The S-matrix was never used and comparison was
not made between Akkerman's approach and the usual wave function approach.
This is done explicitely for the first time in Q1D in this work to
show that the 1D result cannot be extended to Q1D due to the presence
of evanescent modes.

\section{Model and method}

\begin{figure}
\centering{\psfig{figure=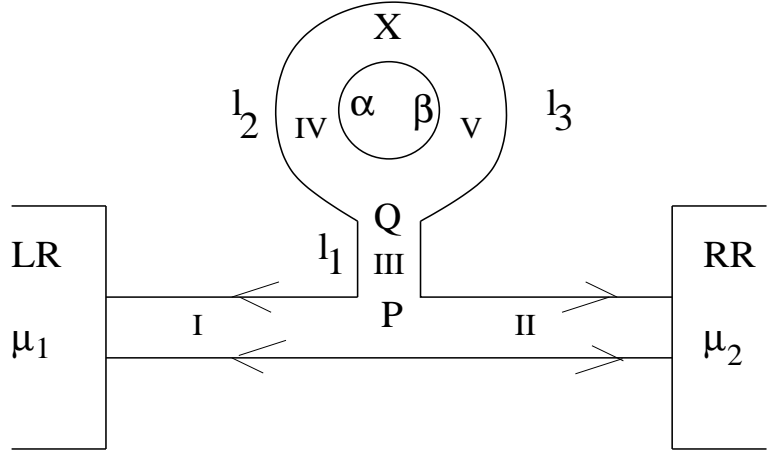,width=10cm,height=6cm,angle=0}}
\caption{A ring connected to an infinite wire.A $\delta$ function potential
is present in the ring at position $X$. A chemical potential
difference ($\mu_1 - \mu_2$) between the left reservoir (LR) and
the right reservoir (RR) drives a transport current through the
regions I and II. The ring is pierced by an Aharonov-Bohm flux
that drives an equilibrium current called persistent current
in the ring.}\label{figsys}
\end{figure}

As shown in Fig. 1, we consider a ring coupled to a wire. 
The scattering solution for this geometry is discussed in detail
in our earlier work \cite{njp}. Here we outline some points
with respect to calculating the RHS of Eq. \ref{ja} which was not done
earlier.
There is a $\delta$-potential impurity
present in the ring at any arbitrary position $X$ [Fig \ref{figsys}]. We apply
Aharonov-Bohm flux $\phi$ through the ring, perpendicularly to the plane 
of the paper.
We consider two modes of propagation because it will show the
shortcomings of Akkerman's formula and that can be
generalized analytically to any number of modes. 
The Schr\"{o}dinger equation for a Q1D wire in presence of a
$\delta$-potential at $x=0$, $y=y_i$ is (the third degree of freedom, i.e.
$z$-direction, is usually frozen by creating a strong quantization \cite{dat})
\begin{eqnarray}
[-\frac{\hbar^2}{2 m^*}(\frac{\partial^2}{\partial x^2}+
\frac{\partial^2}{\partial y^2})+V_c(y)]\Psi(x,y) &=& E\Psi(x,y)
\label{eqse}
\end{eqnarray}
where the $x$-coordinate is along the wire and the $y$-coordinate is
perpendicular to the wire. 
Here $m^*$ is the electron mass and $E$ is the electron energy.
The wave-function in a ring can be obtained
by solving the above equation with periodic boundary condition where
we assume the ring to be so large that its
curvature can be neglected. Here $V_c(y)$ is the confinement potential making
up the quantum wires in Figure \ref{figsys}. The magnetic field just appears as
a phase of $\Psi (x,y)$ that will be accounted for while applying boundary
conditions. Eqn. \ref{eqse} can be separated as
\begin{eqnarray}
-\frac{\hbar^2}{2 m^*} \frac{d^2\psi(x)}{dx^2} &=& \frac{\hbar^2k^2}{2m^*}
\psi(x)
\label{eqsep1}
\end{eqnarray}
\noindent
and
\begin{eqnarray}
[-\frac{\hbar^2}{2 m^*} \frac{d^2}{dy^2}+V_c(y)] \chi_n(y) &=& E_n\chi_n(y)
\label{eqsep2}
\end{eqnarray}
Here we consider that electron is propagating along $x$ direction. This means
in  regions $I$ and $II$ of Fig. 1, $x$ direction is along the arrows. In
region $III$, the $x$ direction is along the line joining P and Q. And in
regions $IV$ and $V$, the $x$ direction is along the perimeter of the ring. One
can choose different axes in the different regions as the matrix equations for
mode matching is independent of this choice \cite{njp}. The
confinement potential $V_c(y)$ in different regions is then in the $y$
(transverse) direction. 
It can be seen from Eqs. \ref{eqsep1} and \ref{eqsep2} that
\begin{equation}
E=E_n+ {\hbar^2 k^2 \over 2 m^*} \label{toten}
\end{equation}
We take $V_c(y)$ to be a square well potential of width $W$, that
gives $\chi_n(y)=\sin [\frac{n\pi}{W}(y+\frac{W}{2})]$. So
$E_n={\hbar^2 n^2 \pi^2 \over 2m^* W^2}$. Hence, in the first mode,
\begin{equation}
k_1=\sqrt{\frac{2 m^*E}{\hbar^2}-\frac{\pi^2}{W^2}} \label{mo1}
\end{equation} 
is the
propagating wave-vector and in the second mode
\begin{equation}
k_2=\sqrt{\frac{2 m^*E}{\hbar^2}-\frac{4 \pi^2}{W^2}} \label{mo2}
\end{equation}
is the
propagating wave-vector.  We have chosen $2m^*=1$ and
$\hbar=1$. 

When electrons are incident along region $I$ (in Fig \ref{figsys})
in the first mode the scattering problem can be solved exactly. The solution to
Eqn. \ref{eqsep1} in region $I$ becomes
\begin{eqnarray}
\psi_I & =& \frac{1}{\sqrt{k_1}}e^{ik_1x}+\frac{r^\prime_{11}}{\sqrt{k_1}} e^{-ik_1x}+\frac{r^\prime_{12}}{\sqrt{k_2}} e^{-ik_2x}
\end{eqnarray}
Similarly, in region $II$, $III$, $IV$ and $V$ we get
\begin{eqnarray}
\psi_{II} & = & \frac{g^\prime_{11}}{\sqrt{k_1}} e^{ik_1x}+ \frac{g^\prime_{12}}{\sqrt{k_2}} e^{ik_2x}\\
\psi_{III} & = & \frac{A e^{ik_1x}}{\sqrt{k_1}} + \frac{B e^{-ik_1x}}{\sqrt{k_1}}+\frac{C e^{ik_2x}}{\sqrt{k_2}} +\frac{D e^{-ik_2x}}{\sqrt{k_2}}\\
\psi_{IV} & = & \frac{E e^{ik_1x}}{\sqrt{k_1}} + \frac{F e^{-ik_1x}}{\sqrt{k_1}}+ \frac{G e^{ik_2x}}{\sqrt{k_2}} + \frac{H e^{-ik_2x}}{\sqrt{k_2}}\\
\psi_V & = & \frac{J e^{ik_1(x-l_2)}}{\sqrt{k_1}} +\frac{K e^{-ik_1(x-l_2)}}{\sqrt{k_1}}+\frac{L e^{ik_2(x-l_2)}}{\sqrt{k_2}} +\frac{M e^{-ik_2(x-l_2)}}{\sqrt{k_2}}
\end{eqnarray}
\noindent
where $r^\prime_{11}$, $r^\prime_{12}$, $g^\prime_{11}$ and $g^\prime_{12}$, 
$A$, $B$, $C$, $D$, $E$, $F$, $G$, $H$, $J$,
$K$, $L$ and $M$ are to
be determined by mode matching. 



\begin{figure}
\centering{\psfig{figure=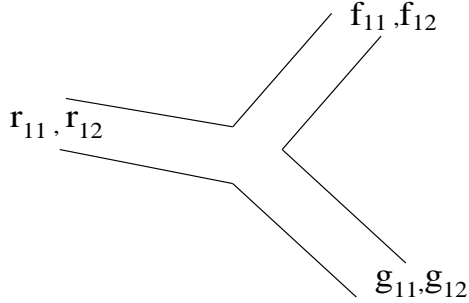,width=6cm,height=4cm,angle=0}}
\caption{A $3$-leg junction.}\label{figleg}
\end{figure}


Note that at $P$ and $Q$ we have a
three legged junction that is schematically shown in Fig. $2$.
In a previous work \cite{njp} we proposed 
a form of junction scattering matrix $S_J$ for a two
channel quantum wire that can be easily generalized to any number of channels.
For the $\delta$ potential impurity at X we use the scattering matrix $S_b$
that was derived by Bagwell \cite{bag}.
One can match the wavefunctions and conserve the currents by
using these $S$-matrices that give us a set of linear euations.
We calculate the coefficients $A$, $B$, $C$, $D$, $E$, $F$, $G$,
$H$, $J$, $K$, $L$ and $M$ numerically by matrix inversion.

Persistent
current can be computed from the wave-function.
\begin{eqnarray}
IW^{(k_1)} &=& \int_{-\frac{W}{2}}^{\frac{W}{2}} \frac{\hbar}{2im^*}
(\psi^\dag\vec{\bigtriangledown}\psi-\psi\vec{\bigtriangledown}\psi^\dag) dy
\label{eqiwf}
\end{eqnarray}
Here the index $k_1$ implies that this is the current due to an incident
electron in $k_1$ channel on the left. Similarly currents are generated
due to incident electron in $k_1$ channel on the right, $k_2$ channel
on the left and $k_2$ channel on the right.
So the net observable persistent current is
\begin{eqnarray}
IW &=& 2 IW^{(k_1)}+2 IW^{(k_2)}
\label{eqiwfk1k2}
\end{eqnarray}
\noindent
From Eq. \ref{eqiwf} we get
\begin{eqnarray}
IW^{(k_1)} &=& 2 I_0 (|E|^2-|F|^2+|G|^2-|H|^2)^{(k_1)}
\label{eqiwfk1}
\end{eqnarray}
\noindent
where $I_0=\frac{\hbar e}{2m^*W^2}$.
\vskip .2in
We also calculate the scattering matrix elements $r^\prime_{11}$,
$r^\prime_{12}, g^\prime_{11}$, $g^\prime_{12}$, $r^\prime_{22}$,
$r^\prime_{21}$, $g^\prime_{22}$, $g^\prime_{21}$ by matrix
inversion to form the scattering matrix of the system as
\begin{equation}
S=\left ( \begin{array}{cccc}
r^\prime_{11} & r^\prime_{12} & g^\prime_{11} & g^\prime_{12}\\
r^\prime_{21} & r^\prime_{22} & g^\prime_{21} & g^\prime_{22}\\
g^\prime_{11} & g^\prime_{12} & r^\prime_{11} & r^\prime_{12}\\
g^\prime_{21} & g^\prime_{22} & r^\prime_{21} & r^\prime_{22}
\end {array} \right ) \label{smsm}
\end{equation}

Substituting \ref{smsm} in \ref{ja},
Eq. \ref{ja} too can be written as a sum of four terms \cite{pra}, 
where each term
consist of scattering matrix elements due to incidence in a particular
momentum channel. That is
\begin{eqnarray}
IS&=& 2 IS^{(k_1)}+ 2 IS^{(k_2)}
\label{eqis}
\end{eqnarray}
where
\begin{eqnarray}
IS^{(k_1)} &=& {1 \over 2 \pi} (\mid r^\prime_{11}\mid^2 \frac{
\partial arg(r^\prime_{11})}{\partial \phi}
+\mid r^\prime_{12}\mid^2 \frac{
\partial arg(r^\prime_{12})}{\partial \phi}+
\mid g^\prime_{11}\mid^2 \frac{
\partial arg(g^\prime_{11})}{\partial \phi}+
\mid g^\prime_{12}\mid^2 \frac{
\partial arg(g^\prime_{12})}{\partial \phi})\nonumber\\
\label{eqisk1}
\end{eqnarray}
and
\begin{eqnarray}
IS^{(k_2)} &=& {1 \over 2 \pi} (\mid r^\prime_{21}\mid^2 \frac{\partial
arg(r^\prime_{21})}{\partial \phi}+\mid r^\prime_{22}\mid^2 \frac{\partial
arg(r^\prime_{22})}{\partial \phi}+\mid g^\prime_{21}\mid^2 \frac{\partial
arg(g^\prime_{21})}{\partial \phi}+\mid g^\prime_{22}\mid^2 \frac{\partial
arg(g^\prime_{22})}{\partial \phi})\nonumber\\
\label{eqisk2}
\end{eqnarray}

Although not implied by the notation, the
currents defined above (Eqs. \ref{eqiwfk1k2} and \ref{eqis}) 
are actually differential currents
in an infinitesimal energy range $dE$. The integration of these
will give the actual measurable currents. The integration
limits depend on the chemical potential $\mu_1$ and $\mu_2$.
Temperature can be included through Fermi function.
All expressions so far is derived for both modes being propagating.
Earlier it was shown that $IW=IS$ for one dimensional ring
coupled to a reservoir \cite{akk}. We shall show below that
when all modes are propagating then one gets $IS=IW$, but not
when we include evanescent modes.
This is because when evanescent modes are present then
some expressions can be analytically continued to include
evanescent modes but not all of them.

\section{Inclusion of Evanescent Modes}

$E$ is the energy of incidence that can be varied as an external
parameter by tuning the chemical potentials of the reservoirs.
When $\pi^2 \le 2m^*EW^2/\hbar^2 < 4\pi^2$, then it can be seen from
Eq. \ref{mo2} and Eqs. 8-12
that $k_2$ mode becomes
evanescent. Eqs. 8-12 are still solutions to Schrodinger Eq. \ref{eqsep1}
implying electrons in the ring can be coupled to an evanescent channel
due to scattering \cite{bag}.
A single impurity can couple an electron to the evanescent second channel.
Scattering at the junctions can also couple an electron to the evanescent
second channel. 
No electron can be incident from (or emitted to)
$\infty$ along the evanescent 2nd
channel.
So the scattering problem has to be solved with an incident
electron in $k_1$ mode in the left lead and an outgoing electron in $k_1$ mode
in the right lead. Hence the $S$ matrix becomes $2\times 2$ and is given by
\begin{eqnarray}
S=\left ( \begin{array}{cc}
r^\prime_{11} & g^\prime_{11}\\ g^\prime_{11} & r^\prime_{11}
\end {array} \right )
\label{eqs}
\end{eqnarray}
Although the $S$ matrix is $2 \times 2$, its calculation has to be done by
using the 6x6 junction S-matrix $S_J$ that is defined in Ref.
[\cite{njp}] and
the 4x4 S-matrix $S_b$ for the $\delta$ function potential that is
defined in Ref. [\cite{bag}].
This is essential because evanescent modes can be coupled inside the
ring without violating any physical principle like conservation of
energy. $g^\prime_{12}$,
$r^\prime_{12}$ etc are also non-zero, but they do not carry any current.
They are not S-matrix elements any more.
Rather they define the coupling to evanescent modes. Unitarity should imply
$\mid r^\prime_{11} \mid^2+\mid g^\prime_{11} \mid^2=1$ and indeed we get
this from the junction matrix defined by $S_J$. This
implies that $S_J$ is appropriate for accounting for realistic multichannel
situations.

\begin{figure}
\centering{\psfig{figure=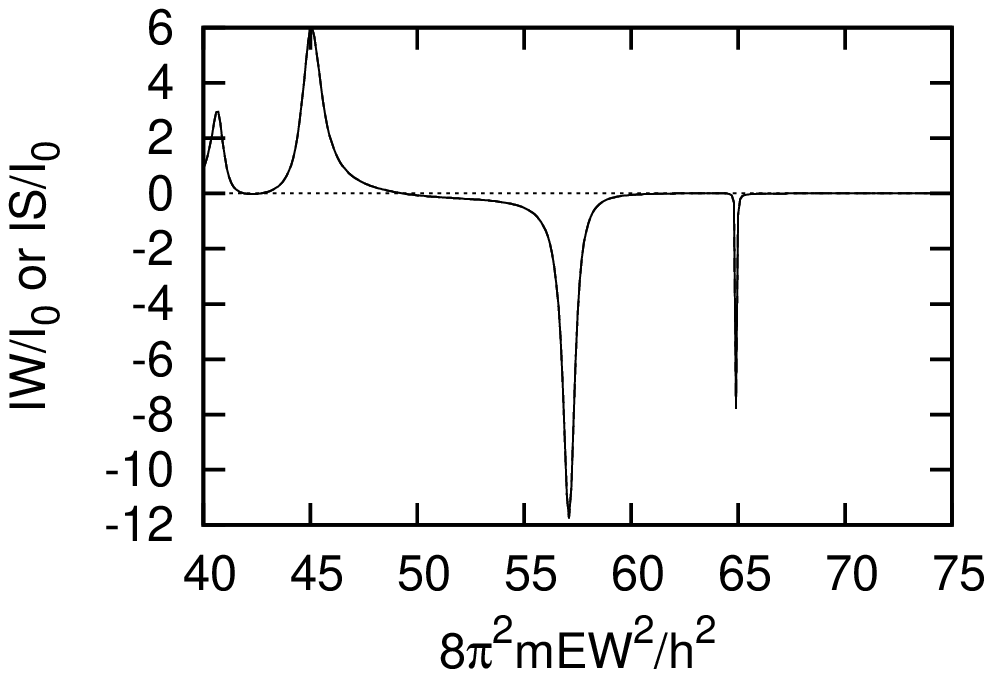,width=10cm,height=6cm,angle=0}}
\caption{$IW/I_0$ and $IS/I_0$ vs $8\pi^2m^*EW^2/h^2$.
The system parameters are
$l_1=l_2=l_3=1$, $y_i=0.1$, $\alpha=\beta=0.3$ and $\gamma=4$.} \label{figprc}
\end{figure}

\section{Results and Discussions}

In a real system, there are always propagating modes as well as evanescent
modes. As is evident from Eqs. \ref{mo1} and \ref{mo2},
evanescent modes have higher transverse energy than the propagating
modes. Also higher the $n$ value of the evanescent mode, higher is its
transverse energy. This energy is at the cost of the propagation
energy which becomes more and more negative for higher $n$
evanescent modes. There will be a natural cut off as very high energies
cannot be realized in a quantum wire. In our simplistic approach we will
first
consider a case when there are two modes in the wire, both being propagating.
We will then consider a case when one mode is propagating and the other
is evanescent. In the first case we will verify that Akkerman's approach
gives exactly the same result as the current calculated from the
wave-function (i.e., $IS=IW$). 
In the second case there are complexities. It will be 
argued that such complexities will persist in a real system where there
will be many evanescent modes.

\begin{figure}
\centering{\psfig{figure=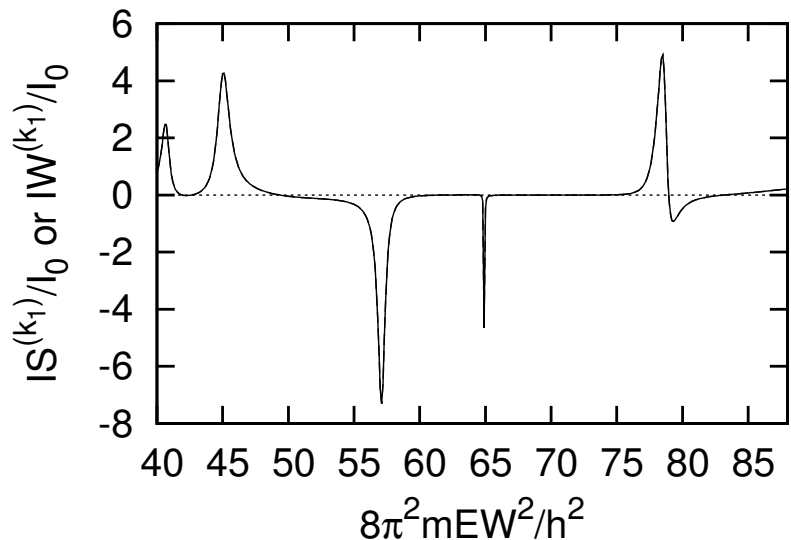,height=8 cm,angle=0}}
\caption{$IS^{(k_1)}/I_0$ and $IW^{(k_1)}/I_0$ versus 
$8\pi^2 m^* EW^2/ h^2$. The system
parameters are $l_1=l_2=l_3=1$, $y_i=0.1$, $\alpha=\beta=0.3$, $\gamma=4$.}
\label{figpik1}
\end{figure}

\begin{figure}
\centering{\psfig{figure=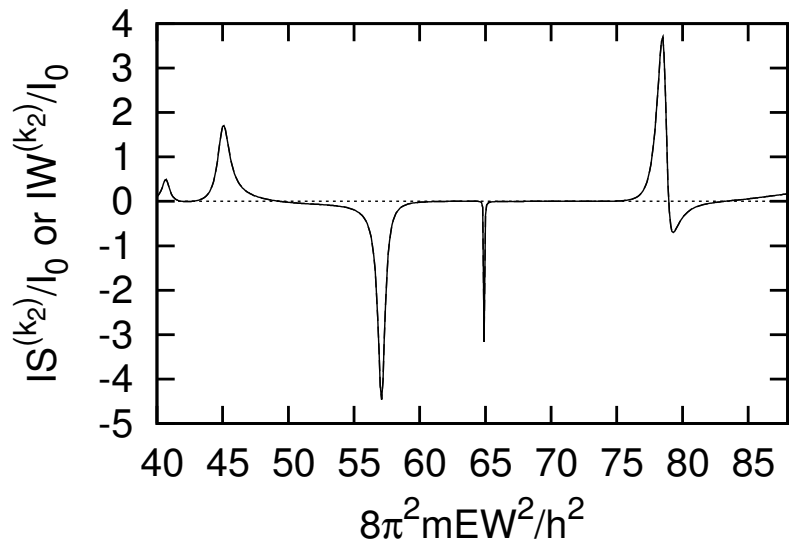,height=8 cm,angle=0}}
\caption{$IS^{(k_2)}/I_0$ and $IW^{(k_2)}/I_0$ versus
$8 \pi^2 m^* EW^2/h^2$. The system
parameters are $l_1=l_2=l_3=1$, $y_i=0.1$, $\alpha=\beta=0.3$, $\gamma=4$.}
\label{figpik2}
\end{figure}

{\it When both modes are propagating:}
First we consider the energy range  $4\pi^2 \le 2m^*EW^2/\hbar^2\le 9\pi^2$
(i.e. $39.478 \le 2m^*EW^2/\hbar^2 \le 88.826$).
Substituting this E in Eqs. \ref{mo1} and \ref{mo2} we can see
that both the modes are
propagating. The nature of current is plotted in Fig \ref{figprc}. The figure
shows that the current $IS$ obtained from Akkerman's formula (that is from
$S$-matrix) is identical with $IW$
(that is from wave function). In fact,
$IS^{(k_1)}$ and $IS^{(k_2)}$ are individually identical with
$IW^{(k_1)}$ and $IW^{(k_2)}$, respectively. These are shown in
Fig. \ref{figpik1} and Fig. \ref{figpik2}. This implies that $IW^{(k_1)}$
is the same algebraic expression as $IS^{(k_1)}$.
Similarly for $IW^{(k_2)}$ and $IS^{(k_2)}$.
And also $IS$ and $IW$ are the same algebraic expressions.


\begin{figure}
\centering{\psfig{figure=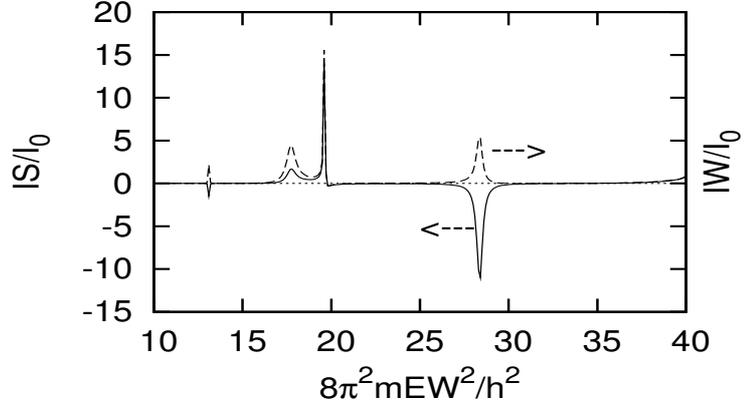,width=10cm,height=6cm,angle=0}}
\caption{$IS/I_0$ (solid line)
and $IW/I_0$ (dashed line) versus $8 \pi^2 m^* EW^2/h^2$ of the
system when second channel is evanescent. The system parameters are
$l_1=l_2=l_3=1$, $y_i=0.1$, $\alpha=\beta=0.3$, $\gamma=4$.
Here $IS/I_0$ is obtained by substituting $S$ given in Eq. \ref{eqs}
into Eq. 1.}\label{figpic}
\end{figure}

{\it When one mode is evanescent:}
Now consider the energy range $\pi^2 \le 2m^*EW^2/\hbar^2 < 4\pi^2$ (i.e.
$9.87 \le 2m^*EW^2/\hbar^2 \le 39.477$)  so that
$k_2=\sqrt{\frac{2m^*E}{\hbar^2}-\frac{4\pi^2}{W^2}}$ becomes imaginary
($k_2\rightarrow i\kappa_2$) while
$k_1=\sqrt{\frac{2m^*E}{\hbar^2}-\frac{\pi^2}{W^2}}$ remains real. In this
regime the ring contains one propagating and one evanescent mode. Evanescent
mode current can be calculated by directly applying Eqn. \ref{eqiwf} to
evanescent mode wave functions or it can be calculated by analytically
continuing propagating mode current 
\ref{eqiwfk1k2} to below the barrier (that is
$k_2\rightarrow i\kappa_2$).
We have already argued 
that $IS$ and $IW$ are the same algebraic expression. 
Under the transformation $k_2\rightarrow i\kappa_2$, applied to both
$IS$ and $IW$, they definitely remain the same algebraic expression.
However, Akkerman's formula takes a different meaning in this regime
where there are evanescent modes.
This is essential because the Akkerman's formula in Eq. \ref{ja} 
is related to the
S-matrix and the transformed expression ($IS_{k_2->i\kappa_2}$) 
cannot be obtained from
the S-matrix.
We know that no electron can be incident along the evanescent channel. So
$IW^{(k_2)}$ and $IS^{(k_2)}$ are zero. 
So $IW$ is now just equal to $IW^{(k_1)}$, where $k_2$ has been analytically
continued. If one assumes that in this regime $IS$
is equal to $IS^{(k_1)}$ where
$k_2$ is analytically continued, then obviously once again $IS=IW$,
as they are the same algebraic expression. However, one can see that
$IS^{(k_1)}$ (see Eq. \ref{eqisk1})
cannot be obtained by substituting the S-matrix (Eq. \ref{eqs}),
into Akkerman's formula (Eq. \ref{ja}). 
If we do this substitution, then we will get the first term 
and the third term in Eq. \ref{eqisk1} (where of course $k_2 \rightarrow 
i \kappa_2$ transformation is taken care of). 
We will not get the 2nd and 4th terms
as $r'_{12}$ and $g'_{12}$ are not S-matrix elements any more.
In fact, we see from
Fig \ref{figpic} that the difference between $IS$ and $IW$ is quite
large.

\begin{figure}
\centering{\psfig{figure=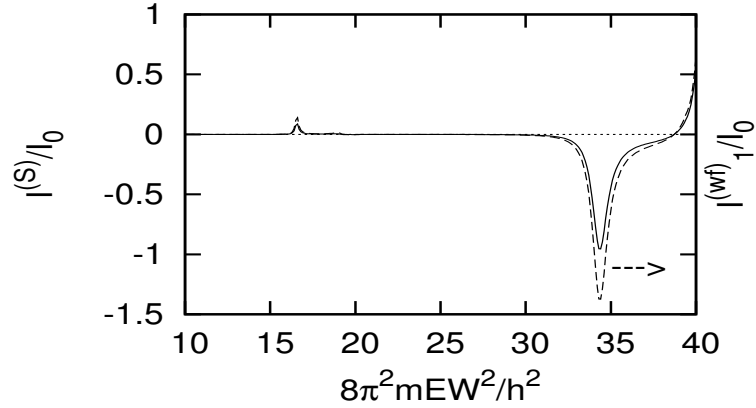,width=10cm,height=6cm,angle=0}}
\caption{$IS/I_0$ (solid line) and $IW^{(k_1)}_1/I_0$ (dashed line) versus
$8\pi^2m^*EW^2/h^2$. The system parameters are $l_1=l_2=l_3=1$, $y_i=0.1$,
$\alpha=\beta=0.3$, $\gamma=-3.7$.
Here $IS/I_0$ is obtained by substituting $S$ given in Eq. \ref{eqs}
into Eq. 1.} \label{figpisi1}
\end{figure}

Since $IS$ is now reduced to just two terms, one may ask the question
that will it give the partial current in the propagating channel only.
Because this partial current also consists of two terms only.
Note from Eqn. \ref{eqiwfk1} that the total current
$IW^{(k_1)}=IW^{(k_1)}_1+IW^{(k_1)}_2$, where
$IW^{(k_1)}_1=2I_0(\mid E\mid^2-\mid F\mid^2)^{(k_1)}$, $E$ and $F$ being
the wave function amplitudes in the propagating channel and
$IW^{(k_1)}_2=2I_0(\mid G\mid^2-\mid H\mid^2)^{(k_1)}$, $G$ and $H$ being
the wave function amplitudes in the evanescent channel. We have plotted 
$IW^{(k_1)}_1$ in Fig. \ref{figpisi1}. The figure shows that $IS$
differs from $IW^{(k_1)}_1$. So it is confirmed that $IS$ obtained
from Eq. \ref{ja} neither give
the total measurable current of the system nor the partial current through
the propagating channel.

\section{Conclusions}

For realistic mesoscopic rings connected to leads, there are always
evanescent modes. The S-matrix is always defined by the
propagating modes only. For such systems one cannot directly apply 
Akkerman's formula. Instead one should start with a model where
all the modes are made propagating. One should apply Akkerman's
formula to the S-matrix of this system and then analytically
continue this expression for the current to the situation
where an appropriate number of modes are evanescent.
While the Landauer-Buttiker approach is still inevitable as
the evanescent modes are obtained due to an incident electron
that is scattered to evanescent modes, the formula given in Eq. 1
is no longer strictly valid in presence of evanescent modes.

\section{acknowledgment}
One of us (PSD) would like to thank ICTP for local hospitality
and other facilities where a part of this work was done.

\end{document}